\documentclass[aps,prl,twocolumn,showpacs,psfig,superscriptaddress,longbibliography]{revtex4-2}
\usepackage{textcomp}
\usepackage{times}
\usepackage{graphicx}
\usepackage{float}
\usepackage{latexsym,amsmath,amssymb,bm,euscript}
\usepackage{color}
\usepackage{subfigure}
\usepackage{epstopdf}
\usepackage[colorlinks=true,linkcolor=blue,citecolor=blue,urlcolor=blue]{hyperref}

\usepackage{soul}
\usepackage[normalem]{ulem}
\usepackage{mathrsfs}
\usepackage{amsmath,amssymb}
\usepackage{lettrine}
\usepackage{xfrac}
\usepackage{xspace}
\usepackage{textcomp}
\usepackage{wasysym}
\usepackage{xr}
\usepackage{lineno}
\graphicspath{{./Figs/}}

\begin{document}

\title{Generic integer and fractional quantum anomalous Hall crystals \\ from interaction-driven band folding}
\author{Hongyu Lu}
\affiliation{Department of Physics and HK Institute of Quantum Science \& Technology,
The University of Hong Kong, Hong Kong, China}
\affiliation{New Cornerstone Science Laboratary, Department of Physics,
	The University of Hong Kong, Hong Kong, China}
\affiliation{State Key Laboratory of Optical Quantum Materials, The University of Hong Kong, Pokfulam Road, Hong Kong SAR, China}
\author{Han-Qing Wu}
\affiliation{Guangdong Provincial Key Laboratory of Magnetoelectric Physics and Devices, School of Physics, Sun Yat-sen University, Guangzhou 510275, China }
\author{Bin-Bin Chen}
\affiliation{Peng Huanwu Collaborative Center for Research and Education, Beihang University, Beijing 100191, China}
\author{Wang Yao}
\affiliation{New Cornerstone Science Laboratary, Department of Physics,
	The University of Hong Kong, Hong Kong, China}
\affiliation{State Key Laboratory of Optical Quantum Materials, The University of Hong Kong, Pokfulam Road, Hong Kong SAR, China}
\author{Zi Yang Meng}
\email{zymeng@hku.hk}
\affiliation{Department of Physics and HK Institute of Quantum Science \& Technology,
The University of Hong Kong, Hong Kong, China}
\affiliation{State Key Laboratory of Optical Quantum Materials, The University of Hong Kong, Pokfulam Road, Hong Kong SAR, China}

\begin{abstract}
Among the extensive studies of  fractional quantum anomalous Hall (FQAH) states,
there recently appears a growing interest in the topological states with coexisting charge density wave (CDW) orders.
Such states are referred to as Hall crystals. 
However, compared to those with integer Hall conductivities, the FQAH crystal (FQAHC) is still elusive even at the level of microscopic model.
In this work, we numerically study a topological flat-band model on triangular lattice with spinless fermions. 
At fractional filling of the Chern band, the nearest-neighbor interaction leads to a commensurate and topologically trivial CDW state. 
Interestingly, the folded mini-band above the CDW gap is non-trivial, and we focus on the doping of it without any projection. A series of (F)QAHC states at (fractional) integer fillings of this mini-band are discovered and some FQAHC state might even exist in less ``ideal'' conditions.
The ground-state degeneracies of such (F)QAHC states are enlarged by the CDW degeneracy and the Hall conductivities -- determined by the fillings of the mini-band -- are different from the fillings of the original Chern band.
We also study the thermodynamics of an FQAHC state and find a compressible CDW phase at intermediate temperatures, which might serve as a precursor of lower temperature FQAHC phase. 
Moreover, we numerically demonstrate that such a generic scheme of doping CDW-folded topological mini-band could be applied to bosonic systems, broadening the platforms of Hall-crystal physics and motivating its exploration in quantum moir\'e and cold-atom systems.

\end{abstract}

\date{\today }
\maketitle

\noindent{\textcolor{blue}{\it Introduction.}---}
Moir\'e systems appear to be promising platforms for studying the strongly-correlated phenomena, including the topological states of matter~\cite{Balents2020_strong_correlation_moire, Yu2019_moire_berry_phase, Li2021_FCI_TMD, Devakul2021_magic_tmd, Mak2022_moire_materials, Wang2024_FCI_mote2, Kwan2024_abelian_FCI_mote2, Lu2024_FCI_composite_fermion}.
Apart from the discovery of fractional quantum anomalous Hall (FQAH) states at fractional fillings of topological bands~\cite{Cai2023_signature_fqah_mote2, Park2023_observation_fqah_mote2, Zeng2023_thermo_evidence_fqah_mote2, Xu2023_Observation_FQAH_tMote2, Lu2024_FQAH_multilayer_graphene}, recent experiments have revealed some exotic states with Hall conductance unequal to the filling of the (Chern) band: $\sigma_\mathrm{H}\neq\nu$ (mostly integer Hall conductance at fractional filling)~\cite{Polshyn2022_topological_CDW, Su2025_TEC_graphene, Lu2025_EQAH_graphene, Waters2025_CI_fractional_filling_pentalayer_graphene}, attracting wide attention~\cite{Patri2024_EQAH_theory, DasSarma2024_thermal_crossover, Huang2024_impurity_crossover_FCI}.
One interesting scenario is that  the Brillouin zone (BZ) is folded due to translation symmetry breaking, leading to the integer filling of the folded band with nontrivial total Chern number~\cite{Lu2025_EQAH_graphene, Waters2025_CI_fractional_filling_pentalayer_graphene, Patri2024_EQAH_theory}. 
Such topological states with coexisting charge-density-wave (CDW) orders are now often referred to as Hall crystals~\cite{Halperin1989_Hall_crystal, Song2024_intertwined_FQAH_CDW, Patri2024_EQAH_theory, Sheng2024_QAHC_mote2, Chen2025_FCI_QAHC_mote2, Soejima2024_AHC, Dong2024_stability_AHC, Zeng2024_AHC, Dong2024_AHC_graphene, Zhou2024_FQAH_multilayer_graphene_moireless, Dong2024_QAH_pentalayer_graphene}. 
However, these Hall crystals mostly exhibit integer Hall conductivities. 
Although there exist theoretical discussions and variational wave-function analysis of FQAH crystal (FQAHC) states~\cite{Song2024_intertwined_FQAH_CDW, Tan2025_variational_FQAHC}, the generic formation mechanism of FQAHC and its microscopic realization from unbiased model computations are yet to be established.

One recent example to realize the FQAHC starts from the symmetric  FQAH state at the same filling, and then breaks the translation symmetry within the topological state by closing the neutral gap while maintaining the charge gap across the transition. In this way, the topological properties are the same with and without the CDW order~\cite{Lu_FQAHS2024, Lu2024_FQAH_FQAHS}.
Besides, considering the widely studied integer QAHC states from band folding~\cite{Pan2022_topological_CDW, Sheng2024_QAHC_mote2, PereaCausin2024_QAHC, Chen2025_FCI_QAHC_mote2}, where the specific topology relies on the existence of the CDW order, it is natural to ask whether and how one can realize the FQAHC states in such band-folding scenario.
The intuitive starting point is to dope holes into the filled topological mini-bands in QAHC states or to dope particles into the lowest unfilled topological mini-bands. But as mentioned above, concrete lattice model computations are yet to be done.
Apart from the ground-state mechanism, the thermodynamics of such FQAHC states is even less explored, which is important for states with intertwined orders~\cite{Fradkin_intertwined_highTc_review} and might provide more information above zero temperature that is of value for experiments.

In addition to quantum mori\'e systems, the ultracold systems on optical lattices are
promising avenues to realize FQ(A)H states~\cite{Sorensen2005_fqh_optical,Aidelsburger2013Hofstadter, Yao2013_FCI_dipolar}, and current realizations are basically based boson~\cite{Julian2023photonFQH, Wang2024_FQH_photon}. However, compared to the fermionic counterpart, the bosonic FQAHC remains uncharted even theoretically.

To address these questions in a controlled and systematic manner, in this work, we study a topological two-band triangular-lattice model with large-scale numerical simulations. 
The nearest-neighbor $V_1$-interaction-driven CDW at $\nu=2/3$ filling of the lower Chern band could triple the unit cell and fold the original BZ. Although the CDW is trivial, we find a $C=-1$ mini-band above the CDW gap.
At full filling of this mini-band, the desired QAHC is stabilized by adding competing interactions, however, the spectrum gap is vanishingly small with only $V_1$, suggesting strong band-mixing with upper bands.
The competing interactions could improve the dispersion and quantum geometry of the mini-band for fractional states, and we find a series of FQAHC states at fractional fillings of this mini-band. With only $V_1$ interaction and less ``ideal'' conditions, we find evidence of just one FQAHC state.
The topological and CDW degeneracies both contribute to the total ground-state degeneracies of the (F)QAHC states, and the Hall conductivity of the FQAHC state is determined by the filling of the mini-band ($\nu^\ast$),  different from the filling of the original band ($\nu$).
Through the thermodynamic study of the $\sigma_\mathrm{H}=-\frac{1}{3}$ FQAHC state ($\frac{e^2}{h}=1$ is taken throughout this paper), we find a compressible CDW phase at the intermediate temperature range,  which might serve as a precursor of lower temperature FQAHC phase. 
Moreover, we demonstrate that such a generic scheme of doping CDW-folded topological mini-band could be applied to bosonic systems, with the observed $\sigma_\mathrm{H}=-\frac{1}{2}$ bosonic FQAHC, broadening the platforms of Hall-crystal physics. 

Based on these results, our work has demonstrated a generic mechanism of (F)QAHC states in both fermionic and bosonic systems, paving the way for further understanding the ground-state and finite-temperature properties of such exotic states of matter, and stimulating the exploration of FQAHC states in quantum moir\'e and cold-atom experiments.

\begin{figure}[htp!]
\centering		
\includegraphics[width=0.5\textwidth]{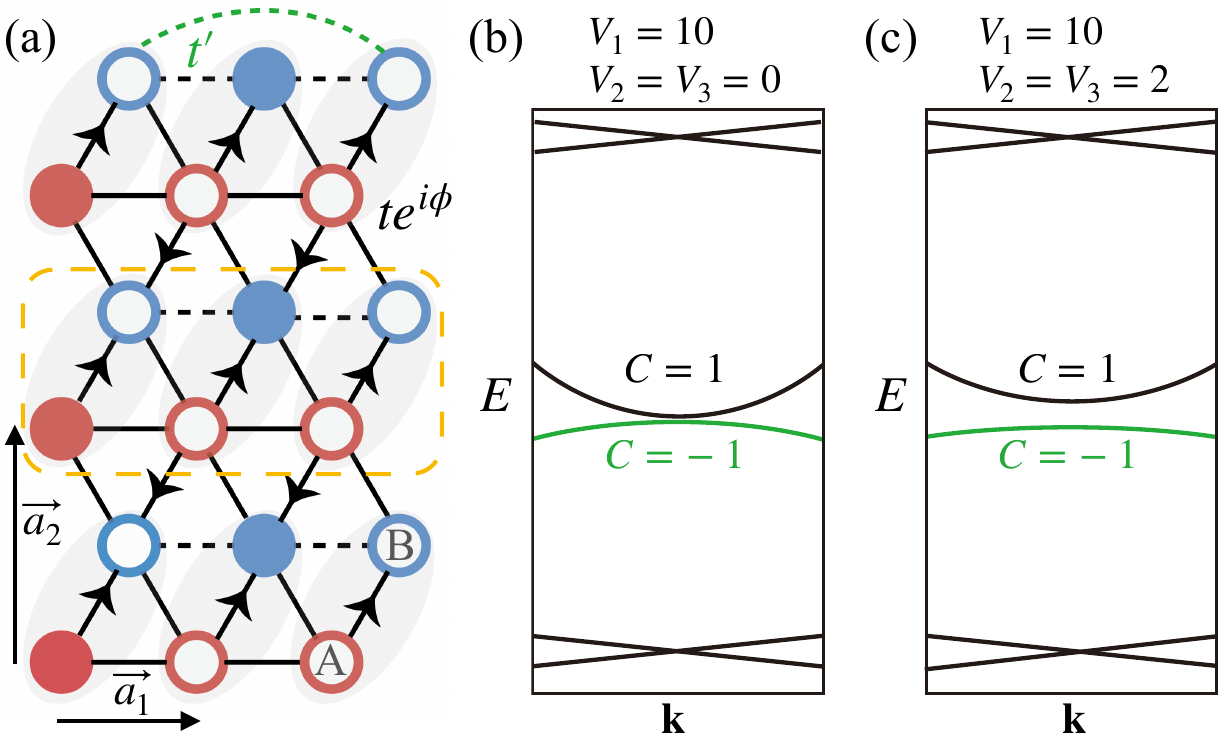}
\caption{\textbf{Model and band folding.} (a) The lattice model with two sites per unit cell and the A/B sublattices are labeled in red/blue. The NN hoppings ($te^{i\phi}$) are labeled by black lines and the phase $\phi$ is 0 on the solid lines, $\pi$ on the dashed lines, $\pi/2$ along the arrows. The green dashed line is an example of the uniform NNNN hopping $t'$.
	The primitive vector of the original two-orbital unit cell is $\vec{a}_1=(1,0)$ and $\vec{a}_2=(0,\sqrt{3})$. The filled/empty sites refers to a pattern of the CDW order (will be discussed later), and the yellow dashed box refers to the corresponding enlarged (tripled) unit cell. 
	The schematic band structures of the folded mini-bands due to the spontaneously translation symmetry breaking are shown  in (b) with only $V_1=10$ and in (c) with $V_1=10,\ V_2=V_3=2$. When there is only $V_1=10$, the spectral gap at $\nu=1$ is vanishingly small [Fig.~\ref{fig_fig2}(a)], suggesting strong band-mixing.
}
\label{fig_fig1}
\end{figure}

\noindent{\textcolor{blue}{\it Model and methods.}---}
We consider a two-band topological model on triangular lattice with the Hamiltonian:
\begin{equation}
	\begin{aligned}
		H =&\sum_{\langle i,j\rangle}te^{i\phi_{ij}}(c_i^\dagger c^{\ }_j+h.c.)+\sum_{\langle\hskip-.5mm\langle\hskip-.5mm\langle i,j \rangle\hskip-.5mm\rangle\hskip-.5mm\rangle}t'(c_i^\dagger c^{\ }_j+h.c.)\\
		&+V_1\sum_{\langle i,j\rangle}n_in_j
        +V_2\sum_{\langle\hskip-.5mm\langle i,j \rangle\hskip-.5mm\rangle}n_in_j
        +V_3\sum_{\langle\hskip-.5mm\langle\hskip-.5mm\langle i,j \rangle\hskip-.5mm\rangle\hskip-.5mm\rangle}n_in_j,
	\end{aligned}
	\label{eq:eq1}
\end{equation}
where the nearest-neighbor (NN) hopping is complex with the fixed magnitude $t=1$ and its phase $\phi_{ij}$ is illustrated in Fig.\ref{fig_fig1} (a).
The next to next nearest-neighbor (NNNN) hopping is uniform and we take $t'=0.2$, which flattens the lower band~\cite{Kourtis2012_FCI_t2g, Kourtis2014_pinball}. 
With these parameters, the Chern number of the upper/lower band is $C=\pm1$, respectively. We will consider the NN, next nearest-neighbor (NNN), and NNNN repulsive interactions $V_1,\ V_2,$ and $V_3$ in this work.

This model has previously been studied~\cite{Kourtis2012_FCI_t2g, Kourtis2014_pinball}, and the so-called topological pinball liquid (TPL) has been reported having coexisting topological and CDW orders and the fractional Hall conductivities are $|\sigma_\mathrm{H}|=\frac{2}{5},\ \frac{3}{5}$ at $\nu=\frac{4}{5},\ \frac{13}{15}$ filling of the Chern band~\cite{Kourtis2014_pinball}.
However, the mechanism of the exotic states have not been adequately revealed. 
The TPL was identified from the real-space perspective in the limit of full polarization, where the fully filled sites of the CDW become localized while the remaining sites construct an effective ``new lattice'' with the rest of electrons filling them.
But the partially filled band of the ``new lattice'' might be topologically trivial and fails to explain the emergent fractional $|\sigma_\mathrm{H}|$~\cite{Kourtis2014_pinball}, unless the kinetic Hamiltonian of the ``new lattice'' is designed to be nontrivial in a different model~\cite{Kourtis2018_newlattice_FCI_CDW}. 
Overall, more clear understanding of the TPL is needed, which is part of the results in our work.

We combine exact diagonalization (ED) and infinite density matrix renormalization group (iDMRG) methods for the ground-state simulations~\cite{McCulloch2008_idmrg}, and the exponential tensor renormalization group (XTRG) method for the finite-temperature simulations~\cite{Chen2018_XTRG}.
In the iDMRG calculations with charge conservation, we consider up to $N_y=6$ unit cells along the periodic direction ($\vec{a}_2$) and mainly focus on $N_x=3$ for the iDMRG unit cells (different $N_x$ have been considered). 
The bond dimension is kept up to $D=3000$ with the maximum truncation error at the  order of $10^{-6}$. 
In the XTRG calculations in the grand canonical ensemble with $H_\mu=\mu\sum_in_i$, we mainly consider the cylinder with $N=3\times18\times2$ sites (different lengths have been tested), and the bond dimension is kept up to $D=500$ with the truncation error $<10^{-4}$ down to very low temperature. In all simulations, we always directly simulate the original Hamiltonian in Eq.~\eqref{eq:eq1} without any projection to any band.

\noindent{\textcolor{blue}{\it CDW and integer QAHC.}---} 
First, let's focus on the $\nu=2/3$ filling of the lower Chern band, which is the electron density $\bar{n}=1/3$. At this filling, large $V_1$ will lead to a commensurate and topologically trivial CDW insulator, whose pattern is illustrated in Fig.\ref{fig_fig1} (a) ($\sqrt{3}\times\sqrt{3}$ for the triangular lattice).  
This CDW would triple the original two-orbital unit cell and thus fold the BZ, leading to effective 6 mini-bands. According to our mean-field analysis at $\nu=2/3$, although the CDW is trivial, the mini-band above the CDW gap has nontrivial topology with $C=-1$, and is well isolated from the two mini-bands below the CDW gap (more details in the Supplemental Information (SI)~\cite{suppl}). 
We define the filling of this $C=-1$ mini-band as: $\nu^\ast=3\times\nu-2$. 
One emergent question is, at half filling of the system, which is the full filling ($\nu^\ast=\nu=1$) of the $C=-1$ mini-band, whether there exist integer QAHC states. If so, there should be 3-fold ground-state degeneracy due to the spontaneously translation symmetry breaking. 
This has not been studied yet.

\begin{figure}[htp!]
	\centering		
	\includegraphics[width=0.5\textwidth]{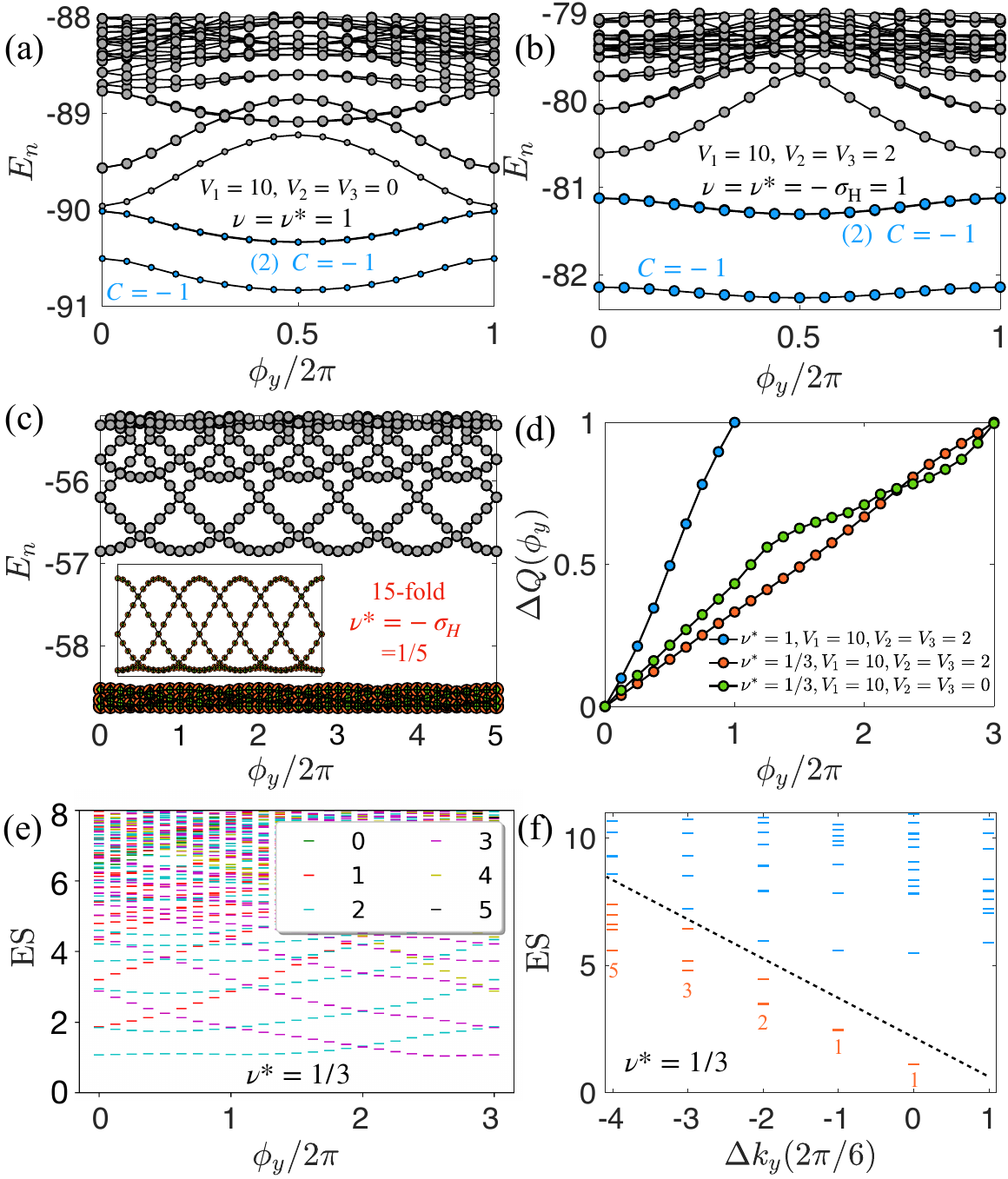}
	\caption{\textbf{Robust (F)QAHC states.} Panels (a-c) are from ED results of a 30-site torus. (a) The spectral flow at $\nu^\ast=\nu=1$ with (a) $V_1=10,\ V_2=V_3=0$ and (b) $V_1=10,\ V_2=V_3=2$. For the QAHC (QAHC$^\ast$ in (a)), there are 3-fold ground states and each with $C=-1$. 
    (c) The spectral flow of the FQAHC state at $\nu=\frac{11}{15}$. There are totally 15 (quasi)degenerate ground states (zoomed in in the inset) and the Hall conductivity from each ground state is $\sigma_\mathrm{H}=\nu^\ast=-1/5$.
    Panels (d-f) are from iDMRG results of $N_y=6$ cylinders.
    (d) Charge pumping of the $\sigma_\mathrm{H}=-1$ QAHC and the $\sigma_\mathrm{H}=-1/3$ FQAHC state with $V_1=10,\ V_2=V_3=2$, and the $\sigma_\mathrm{H}=-1/3$ FQAHC state with $V_1=10,\ V_2=V_3=0$. 
    Panels (e,f) are from $V_1=10,\ V_2=V_3=2$.
    (e) The entanglement spectrum (ES) of the $\sigma_\mathrm{H}=-1/3$ FQAHC state in different charge sectors. After inserting $6\pi$ fluxes, the ES is shifted by one charge sector.
    (f) The momentum-resolved ES of the $\sigma_\mathrm{H}=-1/3$ FQAHC state in the charge sector with the largest eigenvalue, showing characteristic counting of the chiral edge modes of Laughlin states.}
	\label{fig_fig2}
\end{figure}

We take $V_1=10$ as an example. The ED results of a 30-site torus at $V_2=V_3=0$ and at $V_2=V_3=2$ are shown in Fig.~\ref{fig_fig2}(a,b), respectively (we choose $V_2=V_3=2$ for comparison since the CDW gap at $\nu=2/3$ closes around $V_2=V_3\sim4$~\cite{suppl}).
Interestingly, the QAHC state is realized at finite $V_2=V_3=2$ with 3-fold ground states (one from the $\Gamma$ sector and the other two from momenta $(\pm\frac{2\pi}{3},0)$) well gapped from the excited states. Each ground state of this QAHC has $C=-1$, and even the low-energy excited states (identified from their momenta) also have $C=-1$, showing the robustness of this $\sigma_\mathrm{H}=-1$ QAHC. 
This is also confirmed from the iDMRG charge pumping of the $N_y=6$ cylinder in Fig.\ref{fig_fig2} (d)~\cite{Grushin2015_honeycomb_FCI}.
After adiabatically inserting $2\pi$ flux, there is exactly one electron pumped.
We note that, this unique interaction-driven QAHC state appears at the integer filling of the original Chern band, which is incommensurate with the CDW, different from previous QAHCs at fractional fillings (commensurate with CDW) of the original (Chern) bands~\cite{Sheng2024_QAHC_mote2, Chen2025_FCI_QAHC_mote2}.

However, at $V_2=V_3=0$, while the 3-fold states each with $C=-1$ could also be identified, the gap from the lowest excited state is vanishingly small. Although the gap from a 30-site torus is 3 times that of a 24-site~\cite{suppl}, it is still hard to predict its robustness in thermodynamic limit. In this sense, we call it a QAHC$^\ast$, where $^\ast$ suggests the vanishingly small gap.
To understand this, we review the CDW order, and it is obviously also commensurate with the higher filling $\nu=4/3$ (hole version of the $\nu=2/3$ CDW). 
While the $\nu=4/3$ CDW is also trivial, there is another mini-band with $C=1$ below the $\nu=4/3$ CDW~\cite{suppl}, and the higher bands are isolated by the large CDW gap.
Therefore, it is the strong band-mixing between the two mini-bands that leads to the vanishingly small gap of the QAHC$^\ast$ state at $V_2=V_3=0$, but it is different from the ordinary band inversion that leads to trivial topology. Finite $V_2$ and $V_3$ would decrease the band-mixing and stabilize the QAHC. The schematic band structures in the folded BZ are shown in Fig.\ref{fig_fig1} (b-c).

\noindent{\textcolor{blue}{\it Robust FQAHC ground states.}---}
The further question is whether the FQAHC states could be realized if the $C=-1$ mini-band is partially filled.
Naturally, one might expect so at finite $V_2$ and $V_3$ with the stablized QAHC at the integer filling. 
In fact, the previous TPL states in Ref.\cite{Kourtis2014_pinball} with $|\sigma_\mathrm{H}|=\frac{2}{5},\ \frac{3}{5}$ at $\nu=\frac{4}{5},\ \frac{13}{15}$ could be well understood as the FQAHC states at $\nu^\ast=\frac{2}{5},\ \frac{3}{5}$ of the topological mini-band, where they are only realized with finite $V_2$ and $V_3$.

\begin{figure}[htp!]
	\centering		
	\includegraphics[width=0.5\textwidth]{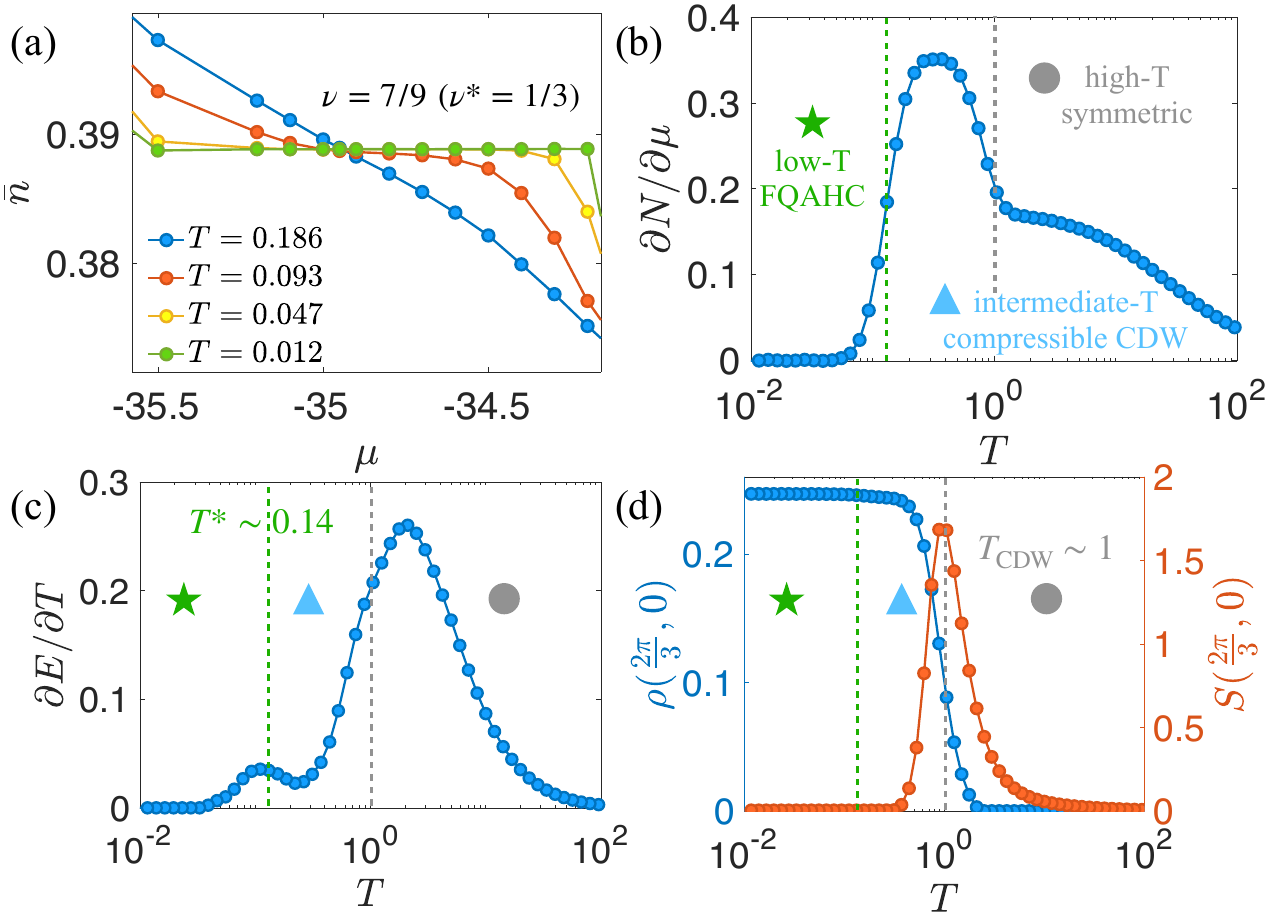}
	\caption{\textbf{Finite-temperature results of the $\sigma_\mathrm{H}=-1/3$ FQAHC state.} Results obtained from XTRG at $V_1=10,\ V_2=V_3=2$: (a) The average density as a function of chemical potential at different temperatures, focusing on the plateaus at $\nu=7/9$.
		The (b) compressibility, (c) specific heat and (d) CDW order $\rho(\frac{2\pi}{3},0)$ and CDW fluctuations $S(\frac{2\pi}{3},0)$ as functions of temperature.
		The green star, blue triangle, and gray circle refer to the low-$T$ FQAHC, intermediate-$T$ compressible CDW, and the high-$T$ symmetric phases, respectively. The green/gray dashed line refers to the onset temperature of the incompressibility and the CDW transition temperature, respectively. }
	\label{fig_fig3}
\end{figure}

Therefore, we start from the simpler case with $V_1=10,\ V_2=V_3=2 $. Apart from confirming the $|\sigma_\mathrm{H}|=\frac{2}{5},\ \frac{3}{5}$ FQAHCs (found by ED), we find a series of new FQAHC states at fractional fillings of this $C=-1$ mini-band and (more details in the SI~\cite{suppl}). We show the spectral flow of a 30-site torus at $\nu=11/15$ in Fig.\ref{fig_fig2} (c) and find there are 15 degenerate ground states and the Hall conductivity (by calculating the many-body Chern number) of each ground state is $\sigma_\mathrm{H}=-\nu^\ast=-1/5$. The topological degeneracy is $5$ while each of them is tripled due to the CDW order, and these ground states flow back to themselves after inserting $10\pi$ fluxes, which further suggests FQAHC state is a Laughlin state with the coexisting CDW order.
However, according to the band-folding scenario, the finite-size effect for ED might be more severe than that for symmetric FQAH states. For example, in the 30-site torus with 15 unit cells, there are only 5 momenta in the $C=-1$ mini-band.
Thus, the robustness of FQAHC states should be examined by larger-scale simulations and we show the $N_y=6$ iDMRG results at $\nu=7/9$ and $V_1=10, \ V_2=V_3=2$ in Fig.\ref{fig_fig2} (d-f). 
After adiabatically inserting $6\pi$ fluxes, one electron is pumped and the entanglement spectrum (ES) is shifted by one charge~\cite{Li2008_ES, Alexandradinata2011_ES}, supporting the fractional Hall conductivity $\sigma_\mathrm{H}=-\nu^\ast=-1/3$.
Further, the momentum-resolved ES shows the characteristic edge-mode counting $\{1,1,2,3,5,...\}$, in agreement with the edge conformal field theory~\cite{Li2008_ES, Cincio2013_topological_order_idmrg, Regnault2015_ES}, which further provides the smoking gun that the topological order in the FQAHC state could coexist with the CDW order via interaction-driven band folding.

The next question is whether there exists FQAHCs with only $V_1$ as the mini-band is topological in principle. 
At $V_1=10$, we only find evidence of the $\sigma_\mathrm{H}=-1/3$ FQAHC at $\nu=7/9$, and the charge pumping result is shown in Fig.~\ref{fig_fig2}(d), while the curve is less straight than that with $V_2=V_3=2$, which might suggest the less uniform many-body Berry curvature~\cite{Sheng2006_chern_number, Lu2022_QAH_dirac}. 
According to the further Hartree-Fock analysis of the $C=-1$ mini-band, compared to the case at $V_2=V_3=0$, the dispersion is flatter and the quantum geometry is more uniform at $V_2=V_3=2$~\cite{suppl}, which might help explain why the other FQAHCs are more easily realized there.

\begin{figure}[htp!]
	\centering		
	\includegraphics[width=0.5\textwidth]{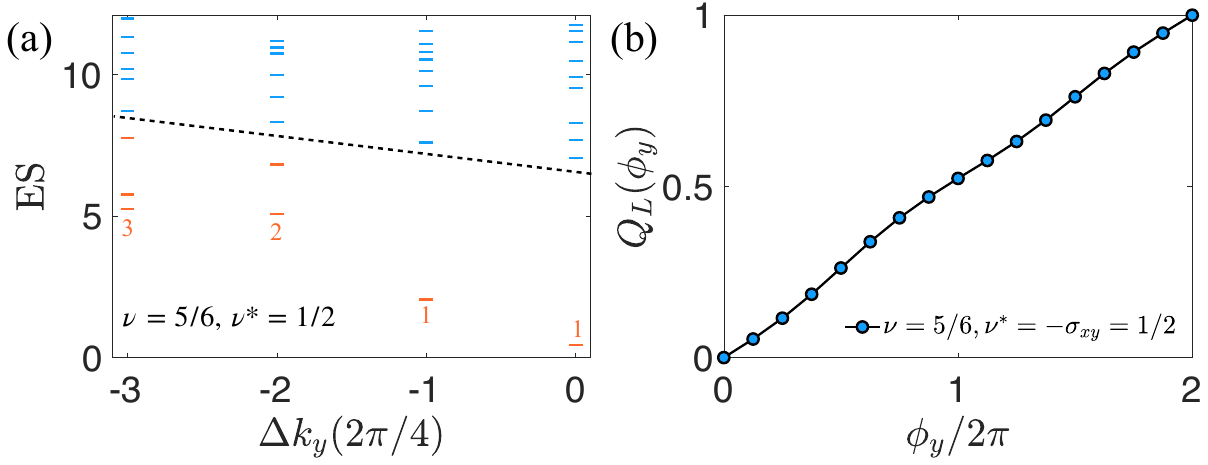}
	\caption{\textbf{Bosonic FQAHC at $\nu=5/6$ at $V_1=10,\ V_2=V_3=0$}.
		(a) The momentum-resolved ES of the $\sigma_\mathrm{H}=-1/2$ FQAHC from iDMRG simulation of the $N_y=4$ infinite cylinder. 
		(b) The charge pumping result of the $\sigma_\mathrm{H}=-1/2$ FQAHC. 
	}
	\label{fig_fig4}
\end{figure}

\noindent {\textcolor{blue}{\it Thermodynamics of the FQAHC state.}---}
Unlike the thermodynamics of symmetric FQAH states~\cite{Lu2024_thermodynamics_FQAH, Lu2024_boson_vestigial}, the finite-temperature knowledge of the FQAHC state with intertwined orders is still elusive, so we showcase the finite temperature properties of the $\sigma_\mathrm{H}=-1/3$ FQAHC state at $V_1=10,\ V_2=V_3=2$.
The $\bar{n}-\mu$ plateaus at $\nu=2\bar{n}=7/9\ (\nu^\ast=1/3)$ are shown in Fig.\ref{fig_fig3}(a), supporting the incompressibility at low temperatures.
The compressibility and specific heat as functions of temperature are shown in Fig.\ref{fig_fig3} (b,c), respectively. We find that the onset temperature of the incompressibility of the FQAHC phase is at $T^\ast\sim0.14$, around which there is a small peak in specific heat and both quantities show exponential decay upon further decreasing temperature.
In addition, the $\bar{n}-\mu$ plateau shrinks with increasing temperatures and is about to vanish above $T^\ast$.
We then further focus on the formation of CDW. The CDW order parameter $\rho(\frac{2\pi}{3},0)$ with $\rho(\mathbf{k})=\frac{1}{N_i}\sum_i e^{i\mathbf{k}\mathbf{r}_i}(\langle n_i^\mathrm{A}\rangle-\bar{n})$ (sublattice $A$ for example and $N_i$ is the number of counted sites), and the structure factor $S(\frac{2\pi}{3},0)$ (referring to such CDW fluctuations) with $S(\mathbf{q})=\sum_{ij}e^{-i\mathbf{q}(\mathbf{r_j}-\mathbf{r_i})}(\langle n_i^\mathrm{A}n_j^\mathrm{A}\rangle -\langle n_i^\mathrm{A}\rangle \langle n_j^\mathrm{A}\rangle)$, are shown in Fig.\ref{fig_fig3} (d).
The CDW fluctuations reach their peak at $T_\mathrm{CDW}\sim1$ (where the CDW order forms), and this CDW transition temperature is higher than the onset temperature of the incompressible FQAHC phase, suggesting the compressible CDW phase at intermediate temperature $T^\ast<T<T_\mathrm{CDW}$, and the symmetric phase at higher temperature $T>T_\mathrm{CDW}$.
Despite the CDW order, the low-energy magnetoroton mode (intrinsic collective excitations in the FQ(A)H states~\cite{GMP1986, Lu2024_thermodynamics_FQAH, Lu2024_boson_vestigial, Zaklama2025_structure_factor_bound, Long2025_spectra_FCI}.) at the different momentum from the CDW vector could still be observed~\cite{suppl}.

\noindent {\textcolor{blue}{\it Bosonic FQAHC state.}---} Furthermore, we show that this generic band-folding scheme for FQAHCs could be applied to bosonic systems. Here, we consider this two-band model filled with hard-core bosons (no more than 1 boson per site) by replacing the fermionic creation/annihilation operators in Eq.~\eqref{eq:eq1} with bosonic ones.
The same trivial CDW at $\nu=2/3$ is observed and we show the iDMRG results at $\nu=5/6$ ($\nu^\ast=1/2$) in Fig.\ref{fig_fig4}. The characteristic counting of edge modes is demonstrated, and after adiabatically inserting $4\pi$ flux, exactly one boson is pumped from right to the left side, supporting the bosonic $\sigma_\mathrm{H}=-1/2$ FQAHC state.

\noindent{\textcolor{blue}{\it Discussions.}---}
In this work, we demonstrate a generic mechanism to realize a series of (F)QAHCs from interaction-driven band folding and renormalization effects, which offers an intriguing combination of the gems of quantum-Hall and Mott physics.
For the FQAHC states here, we find them qualitatively different from the previously found FQAH+CDW state (realized from roton softening in the symmetric FQAH state)~\cite{Lu2024_FQAH_FQAHS}.
The Hall conductivity of the latter is equal to the filling of the Chern band (with $C=1$) and that of the symmetric FQAH state, while the former states have $\sigma_\mathrm{H}=
\nu^\ast\neq\nu$, which might suggest that only part of the electrons in the FQAHC states (from band folding) contribute to the transport while the other part mainly form the CDW.

More importantly, the different roton-driven and band-folding mechanisms together help with the more systematic understanding of the FQAH+CDW states.
We note that these two origins of the FQAH+CDW states are very similar to two mechanisms of realizing supersolids: roton softening in the superfluid~\cite{Chomaz2019_supersolid_dipolar_roton}, and doping the solids with the condensation of the dopants~\cite{Chen2008_SS_dope_solid}. Such analogy of the FQAH+CDW with supersolid physics, further deepens our understanding of these exotic states of quantum matter beyond traditional liquid and solid phases.

While we realize FQAHCs by doping the topological mini-band above the gap of the trivial commensurate CDW, it is interesting to study whether they can be realized by dope the integer QAHC states at commensurate fillings.
Besides, our work suggests the importance of quantum geometry of CDW-driven topological mini-band, which is also mentioned while not explored in recent QAHC experiments~\cite{Su2025_TEC_graphene}. 
To improve the quantum geometry, while we use competing longer-range interactions here, the method might vary in realistic experiments, such as the weak magnetic field for native Chern bands~\cite{Xie2021_topological_CDW_TBG}.

The compressible CDW phase at the intermediate temperature of the FQAHC is also interesting as is might exhibit anisotropic transport properties depending on the specific CDW order, which manifest in the longitudinal resistivity. Reversely, such a finite-temperature CDW metal can be viewed as a precursor of the lower-temperature Hall crystal physics in FQAH systems. Although the metallic CDW phase would not have quantized Hall resistivity, whether there exist nontrivial features of the transverse resistivity at the finite-temperature region of the FQAHC state is interesting for future studies. 

Moreover, we have extended and broadened the platform of Hall crystals to bosonic systems.
Considering the progress made in realizing bosonic FQH states in cold-atom systems,
we believe our work demonstrates a possible scheme to realize FQAHC states in such systems~\cite{Julian2023photonFQH, Wang2024_FQH_photon}.

\begin{acknowledgments}
{\it Acknowledgments}\,---\, We thank Maria Daghofer, Kai Sun, Andrei Bernevig, Nicolas Regnault, Jie Wang, and Yang Liu for helpful discussions. HYL thank Shou-Shu Gong and Xiao-Tian Zhang for discussing the DMRG simulations and Wenqi Yang for discussing the mean-field calculations.
The TeNPy package is used for the infinite DMRG simulations~\cite{tenpy2018} and the QSpace package is used for the thermodynamic simulations~\cite{AW2012_QSpace}. 
HYL and ZYM acknowledge the support from the Research Grants Council (RGC) of Hong Kong (Project Nos. 17309822, HKU C7037-22GF, 17302223, 17301924), the ANR/RGC Joint Research Scheme sponsored by RGC of Hong Kong and French National Research Agency (Project No. A\_HKU703/22). We thank HPC2021 system under the Information Technology Services and the Blackbody HPC system at the Department of Physics, University of Hong Kong, as well as Beijing Paratera Tech Corp., Ltd~\cite{paratera} for providing HPC resources that have contributed to the research results reported within this paper. 
H.Q. Wu acknowledge the support from National Natural Science Foundation of China (No. 12474248), GuangDong Basic and Applied Basic Research Foundation (No. 2023B1515120013), Guangdong Fundamental Research Center for Magnetoelectric Physics (Grant No.2024B0303390001), Guangdong Provincial Key Laboratory of Magnetoelectric Physics and Devices (No. 2022B1212010008). WY acknowledges support by the National Natural Science Foundation of China (No. 12425406), Research Grant Council of Hong Kong (AoE/P-701/20, HKU SRFS21227S05), and New Cornerstone Science Foundation.
\end{acknowledgments}

\clearpage

\section*{Supplementary Information}

\section{A. Band folding analysis}

\begin{figure}[htp!]
	\centering		
	\includegraphics[width=0.5\textwidth]{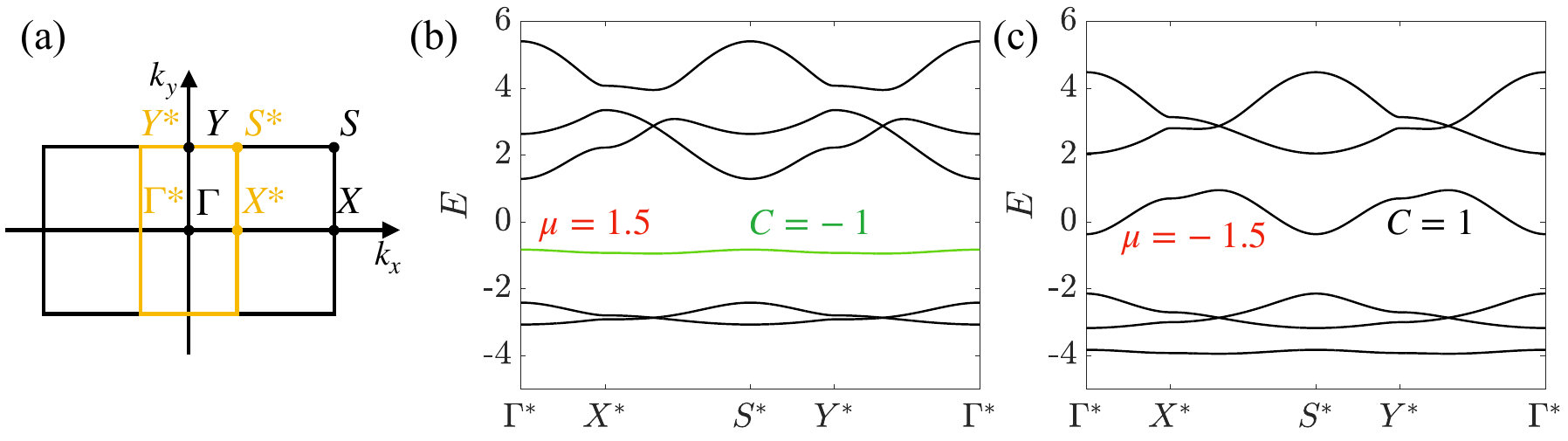}
	\caption{(a) The original BZ and folded BZ ($^\ast$).  The schematic band structures from applying the effective potentials of  (b) the $\nu=2/3$ CDW with $\mu=1.5$ and (c) the $\nu=4/3$ CDW with $\mu=-1.5$.}
	\label{fig_figS1}
\end{figure}

In the main text, the CDW order and the schematic structures of the folded mini-bands are shown.
In fact, this CDW and the band folding could be phenomenologically understood from simple mean-field analysis.
We can add an effective potential with the CDW periodicity to the single-particle part of the origianal Hamiltonian in Eq. 1. 
For this CDW at $\nu=2/3$, the effective potential is $H_\mu=\mu\sum_i(-n_{i,1}+n_{i,2}+n_{i,3}-n_{i,4}+n_{i,5}+n_{i,1})$, where $i$ refers to the enlarged 6-site unit cell and the staggered potential is based on the CDW pattern in Fig. 1(a) of the main text.
The result of taking $\mu=1.5$ is shown in Fig.\ref{fig_figS1}(b). In the folded BZ, we can find a folded mini-band with $C=-1$ above the $\nu-2/3$ CDW gap.
However, this picture does not capture the higher-filling physics above $\nu=1$. Since this CDW is also commensurate with $\nu=4/3$, which shall also open a gap there (which we have verified through many-body simulations).

As the $\nu=4/3$ CDW could be taken as the hole version of the $\nu=2/3$ one, we can take $\mu=-1.5$ for the effective potential and the result is shown in Fig. \ref{fig_figS1}(c). Now we can find another mini-band with $C=1$ below the $\nu=4/3$ CDW gap. Similarly, this picture fails to describe the lower fillings below $\nu=1$. 
The schematic band structures in Fig. 1 of the main text take both the two pictures here into account and are based on the many-body results [Fig. 2(a-b)].

\begin{figure}[htp!]
	\centering		
	\includegraphics[width=0.3\textwidth]{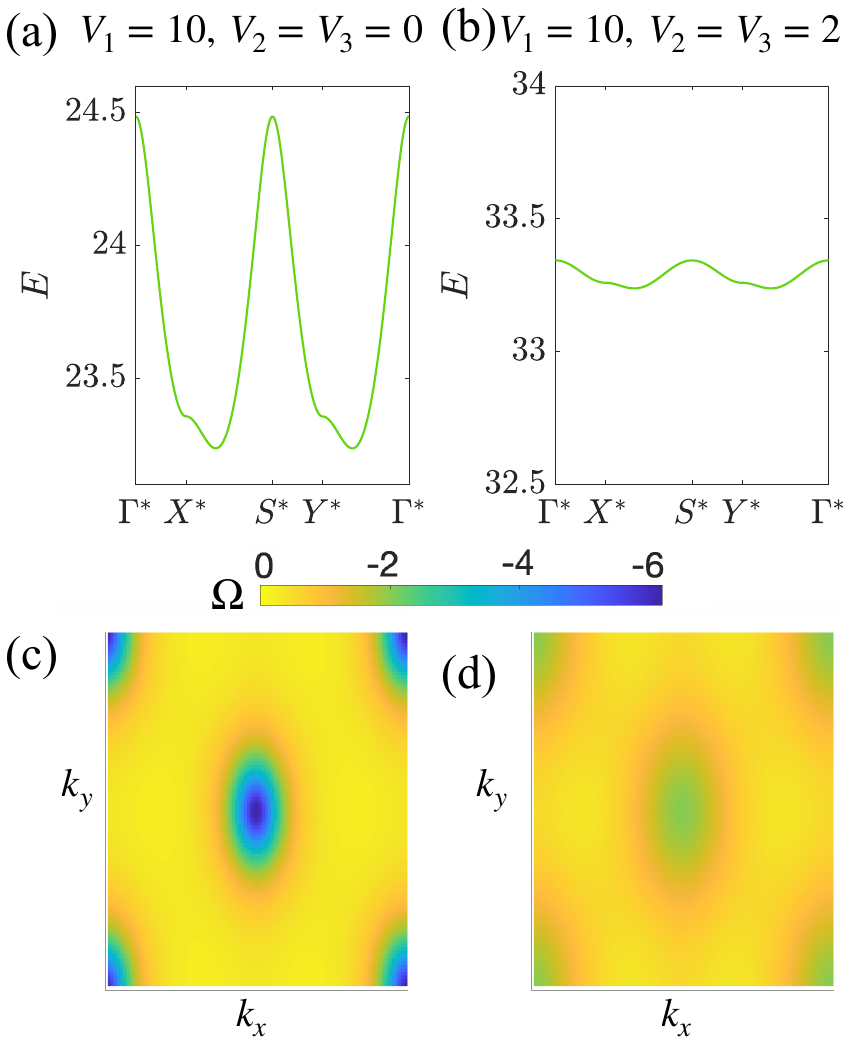}
	\caption{The $C=-1$ mini-band among the interaction-renormalized bands in the folded BZ are
		shown in (a) with only $V_1 = 10$ and in (b) with $V_1 = 10$, $V_2 =
		V_3 = 2$. The bandwidth is $1.25$ and $0.11$, respectively.
		(c,d) The Berry curvature $\Omega$ plotted in the folded BZ
		with the corresponding interactions in panels (a,b) respectively.}
	\label{fig_figS2}
\end{figure}

To help explain the fractional fillings of the $C=-1$ mini-band, more accurate mean-field analysis is required instead of the phenomenological ones above. Therefore, we perform the  Hartree-Fock calculations based on the iDMRG simulations at $\nu=2/3$.
We first show the details of how we obtain the Hartree-Fock (HF) bands renormalized by manybody interactions.

1. The decomposed HF hamiltonian from the original one in Eq. (1) is as follows:
\begin{equation}
	\begin{aligned}
		H_\mathrm{HF} =&\sum_{\langle i,j\rangle}te^{i\phi_{ij}}(c_i^\dagger c^{\ }_j+h.c.)+\sum_{\langle\hskip-.5mm\langle\hskip-.5mm\langle i,j \rangle\hskip-.5mm\rangle\hskip-.5mm\rangle}t'(c_i^\dagger c^{\ }_j+h.c.)\\
		&+V_{ij}\sum_{ij}(\langle n_i\rangle n_j+\langle n_j\rangle n_i-\langle c_i^\dagger c_j^{\ } \rangle c_j^\dagger c_i^{\ }
		-\langle c_j^\dagger c_i^{\ } \rangle c_i^\dagger c_j^{\ }),
		\label{eq:eqs1}
	\end{aligned}
\end{equation}
where we have omitted the constant terms of this single-particle Hamiltonian.

2. For each set of parameters at $\nu=2/3$ filling of the original lower band, we perform iDMRG simulations of the original Hamiltonian (Eq. (1) in the main text) with the same parameters.

3. Then we measure the real-space mean-field parameters in Eq.~\eqref{eq:eqs1} from the many-body wavefunction obtained from iDMRG and put them back to Eq.~\eqref{eq:eqs1}.

4. At last, we diagonalize the 6-band (as our many-body simulations showed that the CDW at $\nu=2/3$ would triple the unit cell) HF hamiltonian with the measured mean-field parameters.

The results are shown in Fig. \ref{fig_figS2}. Here, we only plot the $C=-1$ mini-band above the $\nu=2/3$ CDW gap that we focus on. It is clear that with the finite $V_2=V_3=2$, the bandwidth  of this mini-band is much smaller and the Berry curvature gets more uniform, which provide improved conditions for the FQAHC states at fractional fillings of this mini-band.

\section{B.  The ED spectra at $\nu=2/3$ }
At $\nu=2/3$ with strong $V_1$, the ground state is a CDW. We show the ED spectra of a 24-site torus with fixed $V_1=10$ and tuning $V_2=V_3$ in Fig.~\ref{fig_figS1}. Without competing interactions, the gap of the CDW phase is very large while it gradually and monotonically closes around $V_2=V_3\sim4$. Therefore, we take the intermediate $V_2=V_3=2$ as the example to show the effect of competing interactions  in this work.

\begin{figure}[htp!]
	\centering		
	\includegraphics[width=0.27\textwidth]{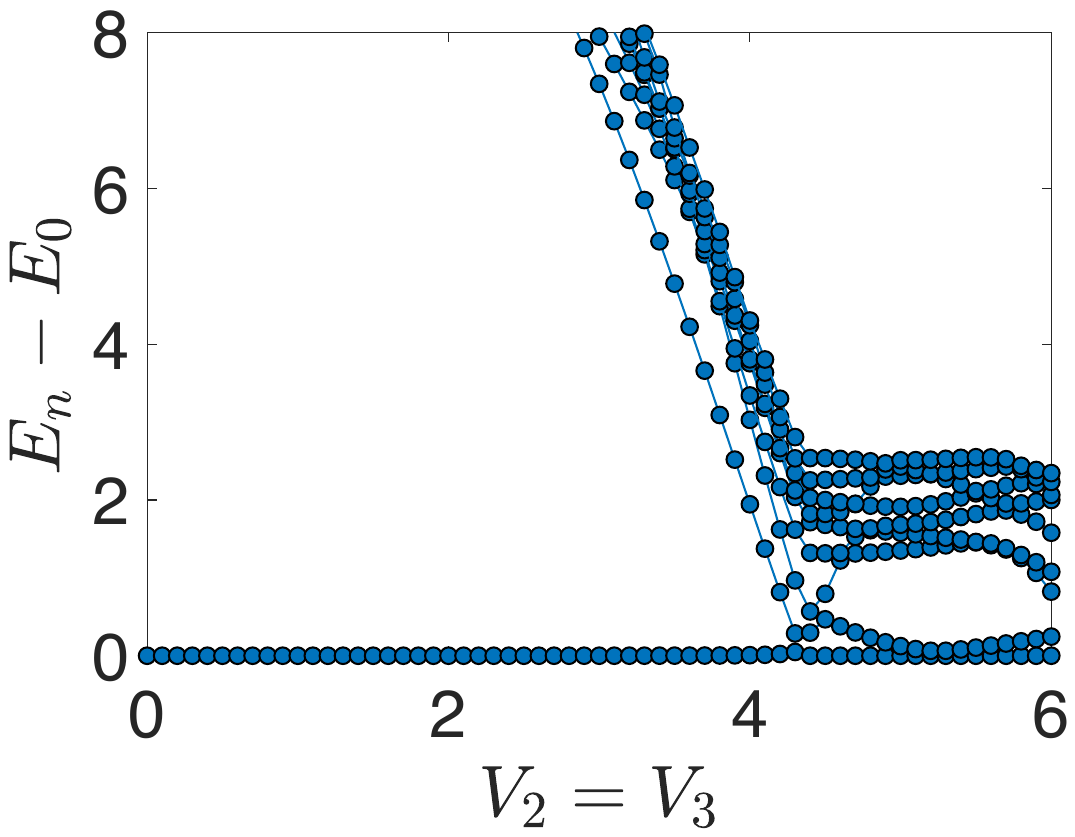}
	\caption{The ED spectra of a 24-site torus at $\nu=2/3$ with fixed $V_1=10$ and tuning $V_2=V_3$. }
	\label{fig_figS3}
\end{figure}

\section{C.  The QAHC$^\ast$ at $\nu=1$ }
\begin{figure}[htp!]
	\centering		
	\includegraphics[width=0.27\textwidth]{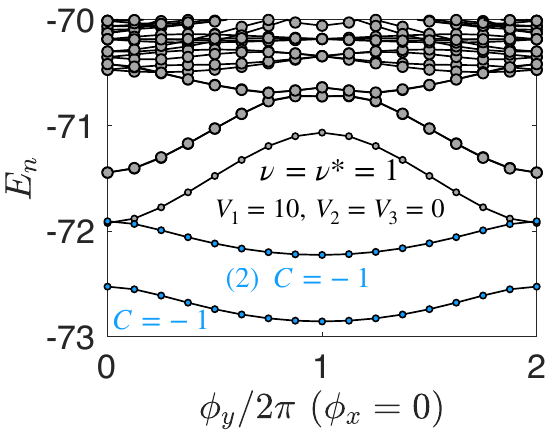}
	\caption{The spectral flow from a 24-site torus at $\nu=1$ with $V_1=10$ and  $V_2=V_3=0$. }
	\label{fig_figS4}
\end{figure}
In the main text, we have shown the ED spectrum of the QAHC$^\ast$ at $\nu=1$ with only $V_1=10$ from a 30-site torus. Here, we show the results from a 24-site torus in Fig.\ref{fig_figS4}.
Similarly, we can identify the lowest 3 states of the QAHC$^\ast$ (the momentum of the fourth lowest state is the same as the lowest state, different from the second and third lowest states), but the gap above the lowest 3 states is vanishingly small. 
For the 30-site case, the gap is $0.053$, and it is $0.017$ for the 24-site case. Although the gap seems to be a bit larger when the system size is larger, it is hard to predict this gap in the thermodynamic limit.

\section{C.  Robust CDW order in the (F)QAHC states } 
\begin{figure}[htp!]
	\centering		
	\includegraphics[width=0.5\textwidth]{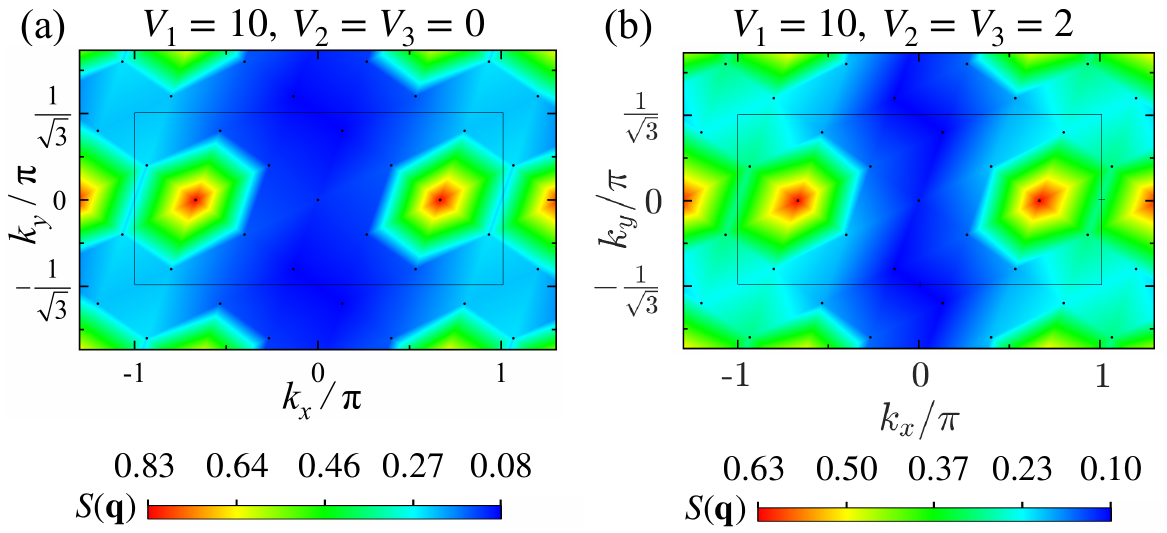}
	\caption{Structure factors from ED results of 30-site tori.}
	\label{fig_figS5}
\end{figure}
In the main text, we have shown the ED spectra at $\nu^\ast=\nu=1$ in Fig. 2(a-b). Here, we show the structure factors of the QAHC($^*$) states in Fig.~\ref{fig_figS5}, respectively, suggesting robust CDW orders. 

\begin{figure}[htp!]
	\centering		
	\includegraphics[width=0.5\textwidth]{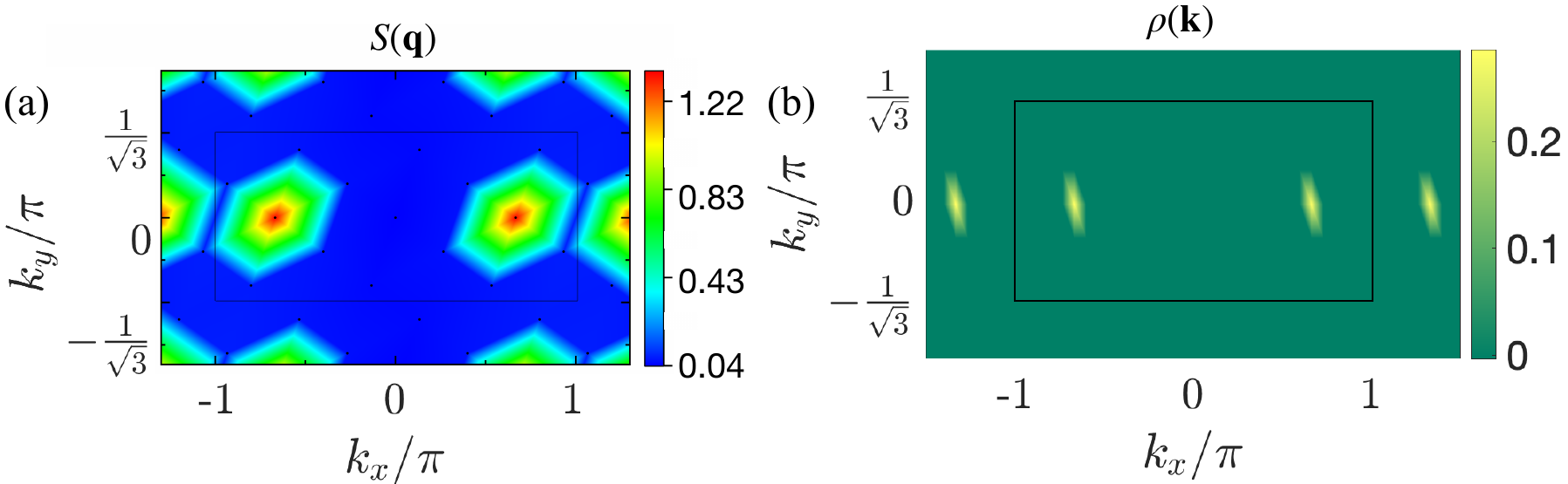}
	\caption{CDW order of the $\sigma_\mathrm{H}=-\nu^\ast=1/5$ state at $\nu=11/15$ with $V_1=10$ and $V_2=V_3=2$. (a) The structure factor  from ED simulations of a 30-site torus.
		(b) The CDW order parameter from DMRG results of a $N_y=6$ cylinder. }
	\label{fig_figS6}
\end{figure}

We further show the robustness of the CDW order in the FQAHC states (taking the $\sigma_\mathrm{H}=-\nu^\ast=1/5$ state at $\nu=11/15$ as an example) in Fig.~\ref{fig_figS6}. The definitions of $S(\mathbf{q})$ and $\rho(\mathbf{k})$ are the same as those in the main text.

\section{D.  Supplementary results of  possible FQAHC states } 
\begin{figure}[htp!]
	\centering		
	\includegraphics[width=0.5\textwidth]{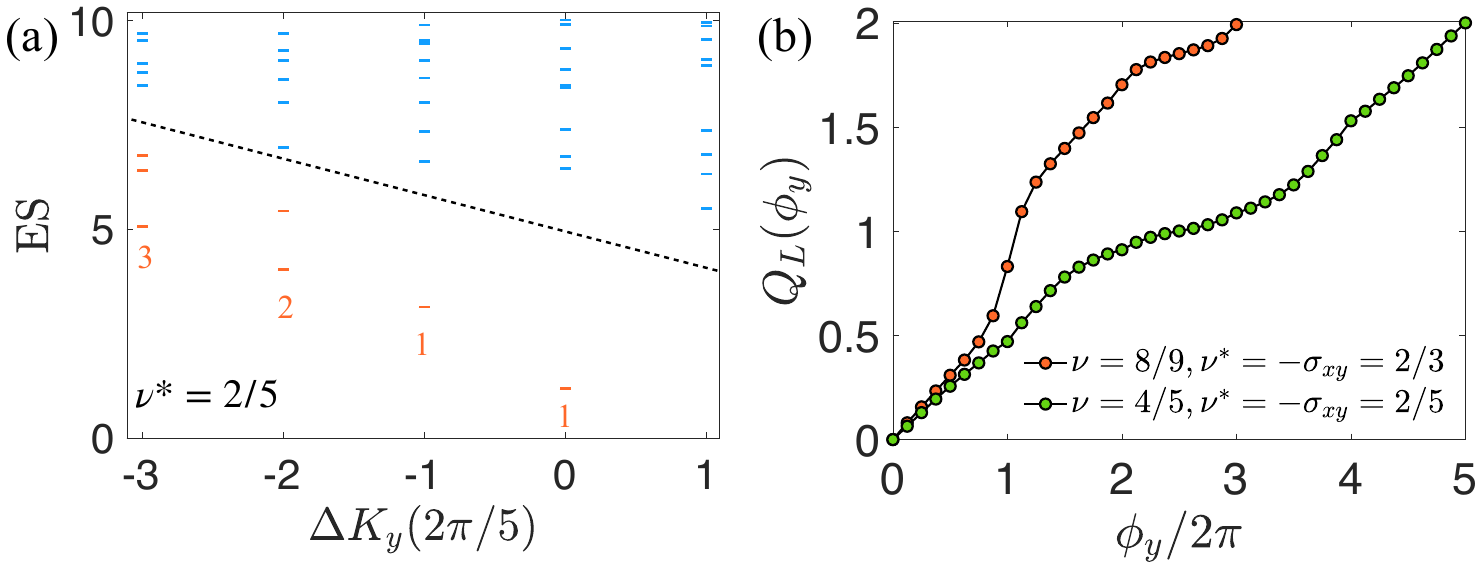}
	\caption{(a) The ES of the $\sigma_\mathrm{H}=-2/5$ FQAHC state at $\nu=4/5$. (b) Charge pumping results of the $\sigma_\mathrm{H}=-2/3$ and $\sigma_\mathrm{H}=-2/5$ FQAHC states. These are obtained at  $V_1=10$ and $V_2=V_3=2$.}
	\label{fig_figS7}
\end{figure}
Here, we first show some more results of different FQAHC states at $V_1=10$ and $V_2=V_3=2$ in Fig.~\ref{fig_figS7}. 
For example, we further show the ES of the $\sigma_\mathrm{H}=-2/5$ FQAHC state at $\nu=4/5$ in Fig.~\ref{fig_figS7} (a), and the counting of states is also consistent with the chiral edge modes of the abelian Jain states.
In addition, we also show the charge pumping results of the $\sigma_\mathrm{H}=-2/3$ and $\sigma_\mathrm{H}=-2/5$ FQAHC states. After inserting $6\pi$ and $10\pi$ fluxes, two quantized fermions are pumped from one edge to the other in both cases. The deviation from straight lines might be related to the non-uniform many-body Berry curvatures or finite-size effect.

\section{Supplementary thermodynamic results}
\begin{figure}[htp!]
	\centering		
	\includegraphics[width=0.5\textwidth]{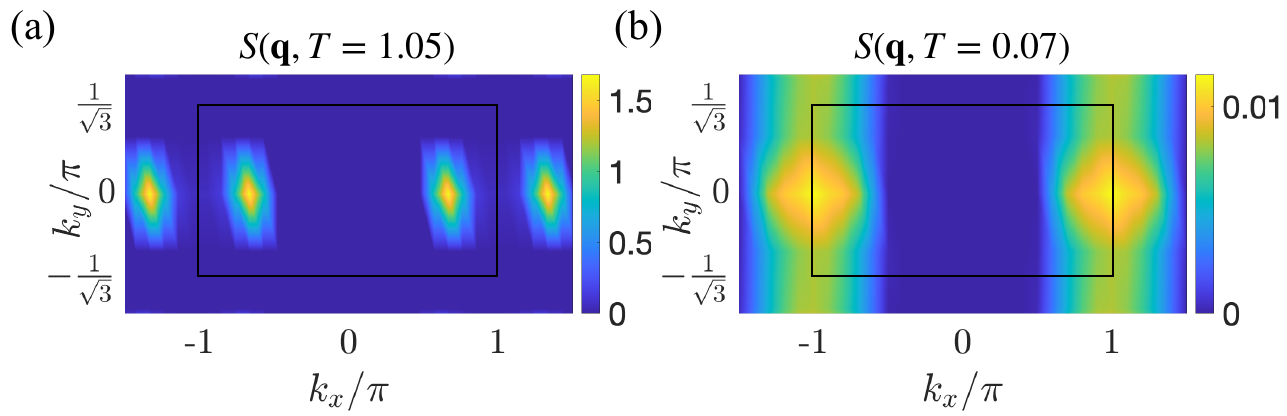}
	\caption{The structure factors in
		the original BZ at different temperatures are shown.}
	\label{fig_figS8}
\end{figure}

We further show the structure factors in the original BZ in Fig.\ref{fig_figS8}(a,b). Around $T_\mathrm{CDW}$, there exist sharp peaks at $\pm(\frac{2\pi}{3},0)$. However, at $T=0.07<T^\ast$, as the CDW order is already well established, we observe that the momenta of the charge fluctuations are no longer at the CDW vectors but around the boundaries of the BZ, which might be related to the low-energy magnetoroton mode (intrinsic collective excitations in the FQ(A)H states).

\clearpage

\end{document}